\documentclass[manuscript]{aastex631}
\usepackage{graphicx}
\usepackage{indentfirst}
\usepackage[utf8]{inputenc}
\usepackage{amsmath}
\usepackage{newtxtext,newtxmath}
\usepackage{bm}

\begin{document}

\title{Dynamics of the Upwind Heliosphere Due to Data-Driven, Solar Wind and Magnetic Field Variations and Implications for Wave Propagation into the Very Local Interstellar Medium}

\author{Chika}
\affiliation{Astronomy Department, Boston University, Boston, MA 02115, USA}

\author{M. Opher}
\affiliation{Astronomy Department, Boston University, Boston, MA 02115, USA}

\author{E. Powell}
\affiliation{Astronomy Department, Boston University, Boston, MA 02115, USA}

\author{S. Du}
\affiliation{Astronomy Department, Boston University, Boston, MA 02115, USA}

\author{J. M. Sok{\'o\l}}
\affiliation{Southwest Research Institute, San Antonio, TX 78238, USA}

\author{J. D. Richardson}
\affiliation{Kavli Institute for Astrophysics and Space Research, Massachusetts Institute of Technology, Cambridge, MA, USA}

\author{B. Van Der Holst}
\affiliation{Astronomy Department, Boston University, Boston, MA 02115, USA}





\begin{abstract}

We introduce an updated, time-dependent treatment to the split-tail (``croissant-like") heliosphere model with data-driven solar wind conditions at 1 au, to study the evolution of the heliosphere with solar-cycle variations in plasma speed, plasma density, and magnetic field intensity. The model produces a sub-Alfvénic and low beta region, not observed by the Voyagers, $\sim$15 au ahead of the heliopause. The simulated magnetic field and radial flow depart from Voyager observations in this region, indicating that time-dependent effects alone are not sufficient to understand this regime of the heliosheath. We decompose fast and slow magnetosonic wave modes from time-dependent plasma pulse structures in the heliosheath, using a linear Riemann variable analysis, for the first time. Fast mode waves can both reflect at the heliopause and transmit into the interstellar medium, and their speeds are unaffected by the low beta plasma regime in front of the heliopause. The model reproduces the pf2 jump in the magnetic field at $\sim$2020 in the interstellar medium and we attribute the source of pressure fronts observed
by Voyager 1 in the interstellar medium, and pressure pulses observed by Voyager 2 in the heliosheath, to fast mode waves that are commonly recurring solar-cycle features. The presence of fast mode waves drive a highly variable termination shock, with average radial speeds of 6.05 au yr$^{-1}$ $\pm$ 5.37 au yr$^{-1}$ in the New Horizons direction. We find that the termination shock has a sinusoidal-like oscillatory motion in the rising phase of the solar cycle, and broad inward motions during the declining phase.


\end{abstract}




\section{Introduction} \label{sec:intro} 

 The heliosphere's interaction with the interstellar medium is a dynamic system, continually shaped by the Sun's changing activity across the solar cycle. Both in-situ and remote observations provide evidence of the solar cycle's influence on the heliosphere and very local interstellar medium. Near the Sun, Ulysses' measurements revealed that at solar minimum the Sun develops large polar coronal holes that emit fast, uniform solar wind plasma at high latitudes while the solar wind at lower latitudes is slower, denser, and more variable \citep{mccomas2000}. During solar maximum, active regions produce highly variable and complex solar wind at all latitudes \citep{mccomas2002}. A comprehensive analysis over several solar cycles from \citealt{sokol2021}, showed the periodic nature of the solar wind dynamic pressure at 1 au, which causes the heliosphere to breathe.

The Voyager spacecraft found that these solar wind variations persist into the outer heliosphere. Voyager 1 (V1) and 2 (V2) also showed the importance solar cycle effects in the heliosheath where the magnetic field intensity, plasma density and temperature, and energetic particle intensity showed correlated fluctuations that can last several months \citep{richardson2009,richburlaga2013,richardson2017, burlaga2021a}. The solar cycle also affects the very local interstellar medium, as inferred from radio emissions, measured every solar cycle since 1980 by V1 \citep{gurnett1993, gurnettkirth2019}. Radio emission and plasma oscillation events were observed for more than 20 au past the heliopause crossing of V1 at 121 au and are thought to be associated with outward moving shocks \citep{burlaga2021,kurth2023}. To complement Voyager in-situ observations, line-of-sight energetic neutral atom (ENA) observations provide a global image of the heliosphere and plasma processes at the boundary of the heliosphere and interstellar medium. Elevated ENA fluxes following periods of high solar wind dynamic pressure indicate their sensitivity to the solar cycle, and also the time-varying characteristics of the solar wind plasma from which they originate \citep{dialynas2017a,dialynas2017b, mccomas2020, sokol2021, mccomas2024}. 

Time-dependent modeling has advanced our interpretation of observational data by simulating solar-cycle variations in the solar wind and showing how these variations propagate into the outer heliosphere. Several studies demonstrated how time-varying solar wind pressure drives periodic motions in the termination shock and heliopause \citep{wangbelcher1999,zankmuller2003, izmodenov2008,izmo&alex2020, washimi2011, washimi2012, provornikova2014}. Other models showed how the solar cycle leaves imprints in the heliosheath, where these effects can produce large-scale turbulence and fluctuations in the heliosphere, upwind (in the Sun's direction of motion) and downwind (opposite of the Sun's direction of motion) \citep{scherer2003,pogorelov2015,washimi2011,washimi2017,fraternale2024}. More recently, time-dependent models extend their analysis past the heliopause to investigate the arrival of solar transients in the interstellar medium \citep{kim2017, pogorelov2021, zirnstein2024}. 

The split-tail, or “croissant-like”, model of the heliosphere has been extensively explored with steady-state solar wind conditions, a regime in which variations across the heliosphere can be attributed solely to spatial effects \citep{opher2015,opher2020,opher2021}. Here we present an updated, time-dependent and data-driven MHD model of the heliosphere that includes varying solar wind conditions over the 11-year solar activity cycle. We study the spatiotemporal effects arising from solar cycle variations in the upwind region of the heliosphere and interstellar medium. Section \ref{sec:2.1} details the components utilized in our model and our implementation of time-dependent boundary conditions that capture solar cycle activity. In section \ref{sec:3} we show model results, focusing on the plasma and magnetic field conditions in the heliosheath and interstellar medium, as well as the motion of the heliospheric boundaries. Furthermore, we place our findings in the context of observations from Voyager and predictions for New Horizons (NH). Section \ref{sec:4} summarizes the main outcomes of the time-dependent model and discusses physical phenomena related to time-dependent effects.

\section{Model and Boundary Conditions} \label{sec:2.1}

 We use the Block-Adaptive Tree Solar wind Roe-Type Upwind Scheme (BATS-R-US) MHD solver of the Space Weather Modeling Framework (SWMF) to evolve a 3D time-dependent, multi-fluid model of the solar wind and interstellar medium interaction \citep{opher2003, toth2012}. The model simulates charge exchange which includes neutral hydrogen and ions. We use a single-ion model where pickup ions and thermal solar wind ions are evolved as a single fluid, as opposed to a multi-ion model where pickup ions and thermal solar wind are treated separately. As in \citealt{opher2009} we obtain the neutral solution using a multi-fluid treatment. We treat neutrals as four fluids, corresponding to four distinct regions of the heliosphere and interstellar medium.The model is capable of describing neutral hydrogen atoms kinetically \citep{michael2022,chen2024}, but a multi-fluid approach is computationally less expensive and the kinetic solution only produces slight differences.

The MHD model includes static mesh refinement on a Cartesian, Sun-centered grid at (x, y, z) = (0, 0, 0) with dimensions -1500 au $<$ x $<$ 1500 au and -2000 au $<$ y,z $<$ 2000 au, and spherical inner boundary at 1 au. The X axis is defined by the line formed at the intersection of the solar equatorial and ecliptic plane, the Z axis is oriented along the solar rotation axis, and the Y axis completes the right-handed coordinate system. Figure \ref{fig:fig2}a shows the model's grid resolution up to 1 au refinement. The upwind region, comprised by 1 au sized grid cells, captures the termination shock, heliosheath, heliopause, and local interstellar medium which are the main regions of focus for this work. The highest level of refinement of 0.125 au (not shown) is placed around the inner boundary, where the solar wind is initialized, to accurately resolve time and latitude varying plasma conditions at the inner boundary. The outer boundary is the plane x= -1500 at the edge of the simulation box, where the interstellar medium flows inward toward the heliosphere with the following ion and neutral hydrogen properties: $v_{ISM}$ = 26.3 km s$^{-1}$, $T_{ISM}$ = 6519 K, $n_{ISM}$ = 0.06 cm$^{-3}$, $n_{H}$ = 0.18 cm$^{-3}$ (same as Case B in \citealt{opher2020}). The neutral and ion components of the interstellar medium have the same velocity and temperature. We take the ISM flow direction at an ecliptic longitude of 75.4$^\circ$ and latitude of -5.2$^\circ$. The interstellar magnetic field, $\textbf{B}_{ISM}$, in the model has a strength of 3.2 $\mu$G, directed at an ecliptic longitude of 47.3$^\circ$ and latitude of -34.62$^\circ$ \citep{zirnstein2016,opher2020}.

\subsection{Time-Dependent Boundary Conditions}\label{sec:2.2}

\begin{figure}[ht!]
    \centering
    \includegraphics[width=1\linewidth]{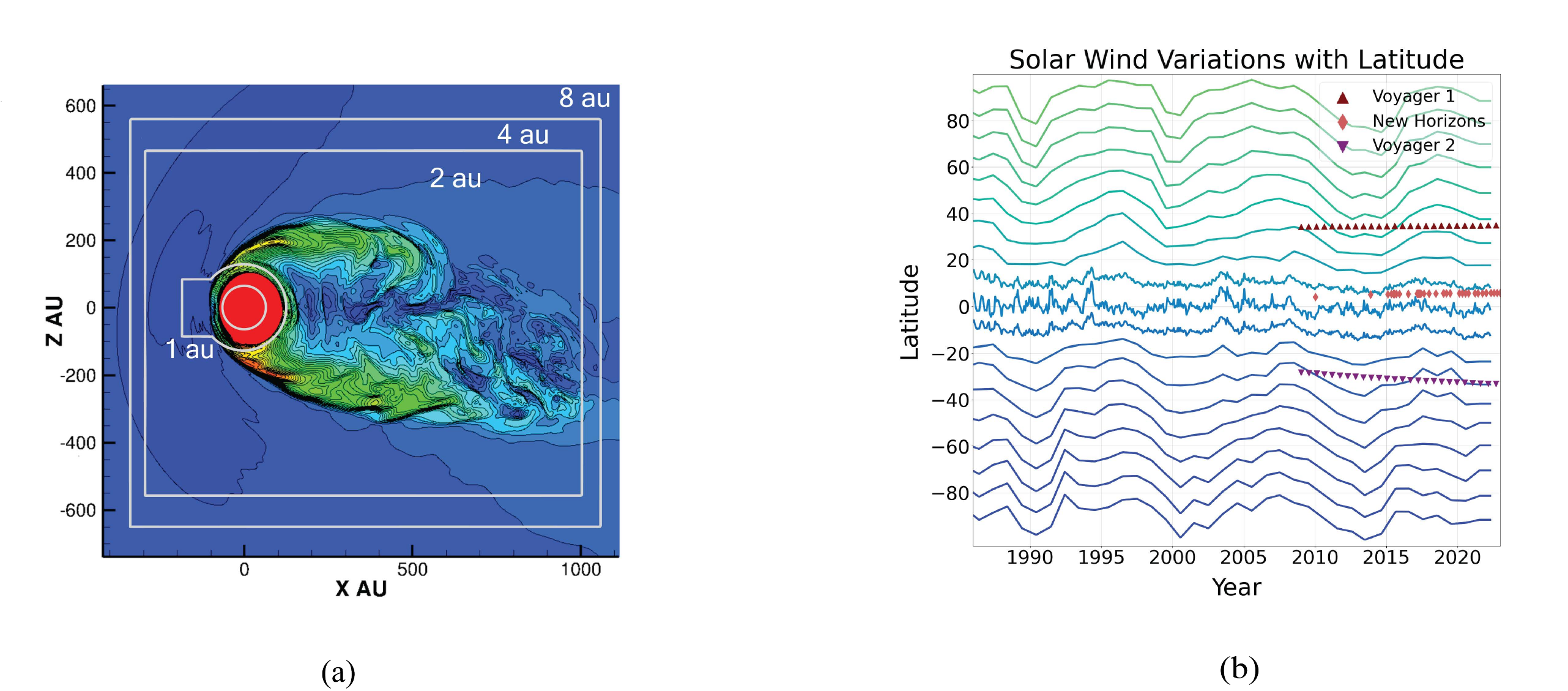}
    \caption{(a) Time-dependent simulation snapshot showing the different levels of refinement in a meridional slice. The background plot shows plasma speed. (b) Variations of the solar wind speed in the model have higher temporal resolution at lower latitudes, intersecting with the New Horizons trajectory. Mid to high latitude regions have lower resolutions, but still capture yearly trends.}
    \label{fig:fig2}
\end{figure}

\begin{figure}[ht!]
    \centering
    \includegraphics[width=1\linewidth]{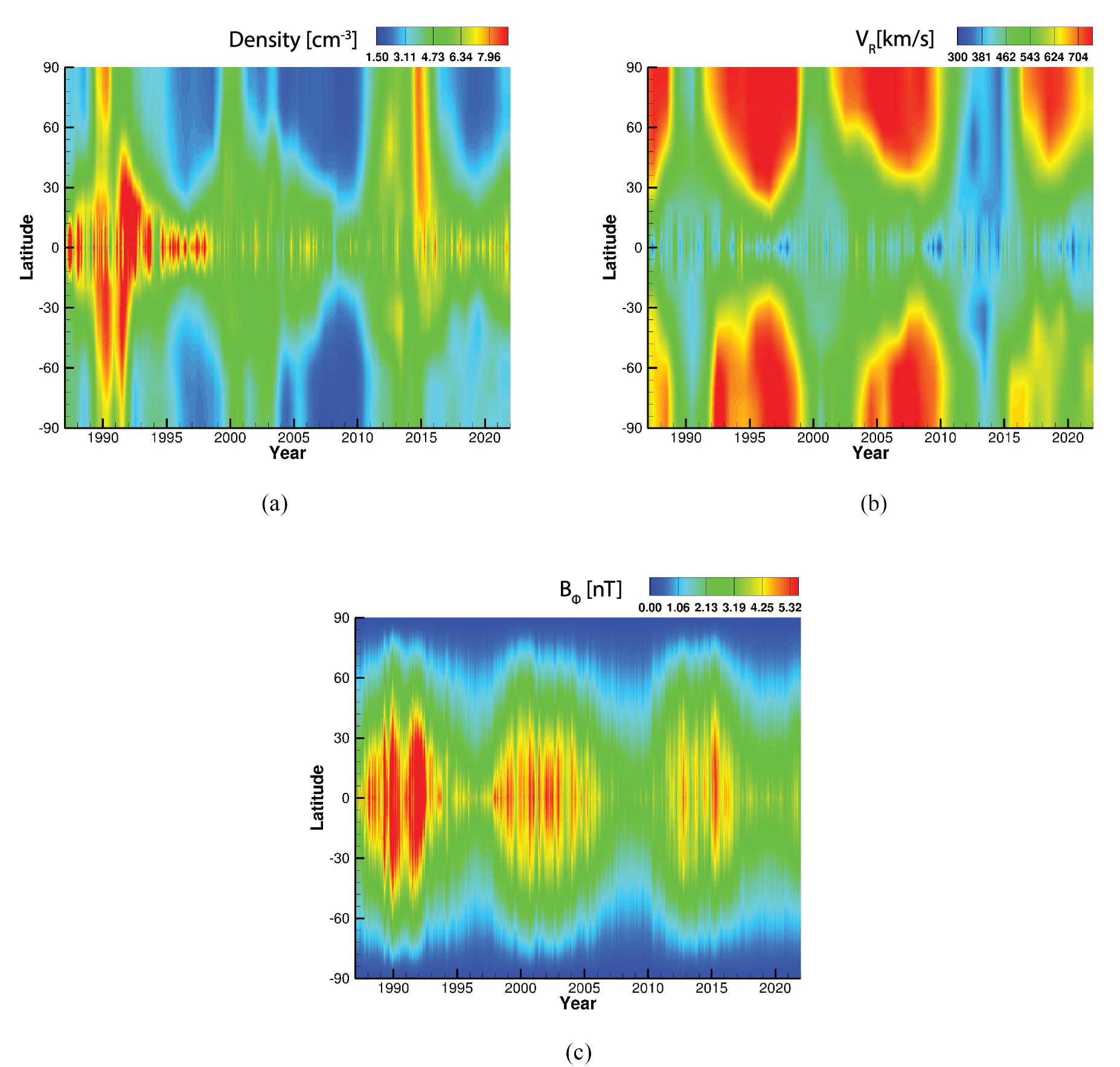}
    \caption{Inner boundary conditions for the solar wind at 1 au following the \citealt{sokol2020} model. Time-dependent (a) plasma proton density, (b) plasma speed, and (c) magnetic field strength (from OMNI data) are the time and latitude-varying parameters input into the model. The time-dependent variations in plasma speed provide the latitudinal gradients in the magnetic field based on the azimuthal component: $\mathbf{B_{\phi}} = B_{0} \left(\frac{R_{0}}{r}\right)^{2}\left(\frac{\Omega \:sin\theta\:r}{u_{sw}} \boldsymbol{\hat{\phi}} \right)$. }
    \label{fig:fig1}
\end{figure}

We use 2D solar wind speed and density profiles from \citealt{sokol2020}, derived from ground-based observations of the latitudinal variations of the solar wind speed through the interplanetary scintillations (IPS) \citep{tokumaru2010,tokumaru2012}, as model inputs. The profiles give time and latitude-varying solar wind conditions at 1 au for solar cycles 22-24 (1985-2022). We calculate the solar wind plasma temperature variations at 1 au using the solar wind speed of the time-dependent data set and a fixed 22-year averaged, near-Earth sonic Mach number of \emph{M} = 6.44 \citep{izmo&alex2020,kornbleuth2021b}. The solar wind input is interpolated and averaged over Carrington rotation, giving the time-dependent model $\approx$27 day resolution, and organized into 10$^\circ$ heliolatitudinal bins from -90$^\circ$ to +90$^\circ$. We assume the time and latitude structure of the solar wind is uniform with longitude to complete a 3D heliospheric structure. The solar cycle trends in the solar wind at the inner boundary are shown in Figure \ref{fig:fig1}a, b. Predominant solar cycle features such as the latitudinal gradients during solar minimum can be seen in the plasma speed and plasma density. We note the decrease in temporal resolution of the solar wind data set, outside of the solar equatorial plane ($>$0$^\circ$ latitude) in Figure \ref{fig:fig2}b. Figure \ref{fig:fig2}b shows the plasma speed, scaled by an arbitrary value to be centered at each 10$^\circ$ latitude bin. It reflects the difference in temporal resolution of solar wind speed data from IPS observations out of the ecliptic plane in comparison to highly resolved OMNI data used to inform the conditions in the ecliptic plane \citep{OMNI,sokol2020}. 

The importance of varying the solar magnetic field intensity with time was shown previously in \citealt{michael2015}, which found that changing the magnetic field improved their model's agreement with Voyager 2 magnetic field measurements in the heliosheath. We incorporate a time dependent solar magnetic field at 1 au, inputting 27-day averages of the field magnitude from 1985-2022, from OMNI data in the ecliptic plane \citep{OMNI}. The solar magnetic field azimuth angle is set to 45$^\circ$ which is the average Parker spiral angle at 1 au. The azimuthal component of the magnetic field, $\mathbf{B_{\phi}} = B_{0} \left(\frac{R_{0}}{r}\right)^{2}\left(\frac{\Omega \:sin\theta\:r}{u_{sw}} \boldsymbol{\hat{\phi}} \right)$, provides the changing latitudinal structure of the field (Figure \ref{fig:fig1}c) due to its dependence on the solar wind speed which varies as described previously. We use a unipolar configuration for the solar magnetic field and neglect the tilt of the solar magnetic axis with respect to the solar rotation axis. These two conditions minimize artificial magnetic dissipation effects that would occur where the magnetic field changes polarity (Opher et al. \citeyear{opher2015}; Izmodenov \& Alexashov \citeyear{izmod2015}; Michael et al. \citeyear{michael2018}).

\begin{figure}[ht!]
    \centering
    \includegraphics[width=1\linewidth]{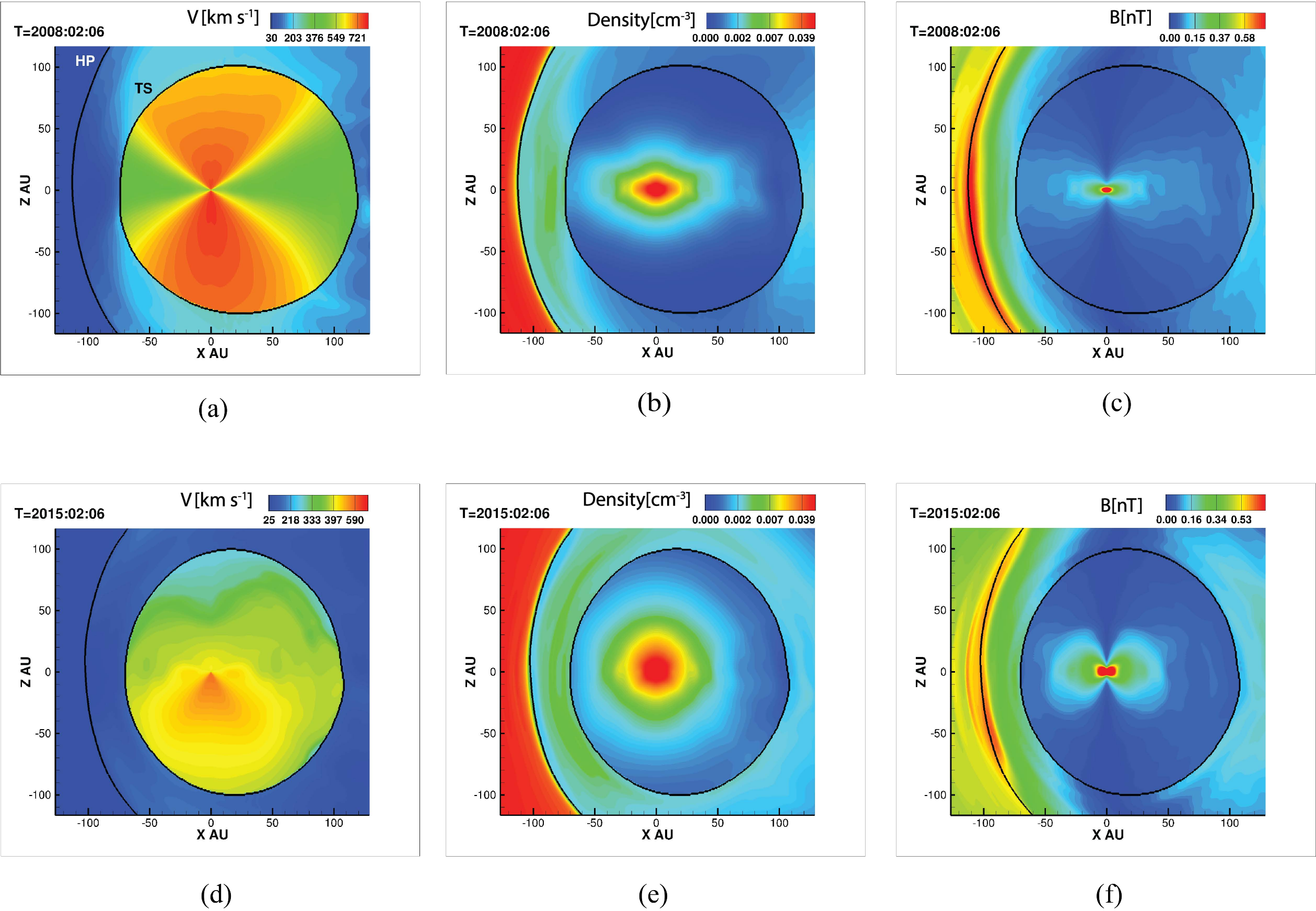}
    \caption{Meridional cuts from the MHD model showing solar cycle characteristics in the solar wind plasma and magnetic field at snapshots of solar minimum (top row, 2008) and maximum (bottom row, 2015) periods. Changes in the (a, d) plasma speed, (b, e) density, and (c, f) magnetic field strength injected at the inner boundary fill up the supersonic solar wind region. (a) Two bold black lines show the location of the termination shock (TS) and heliopause (HP) in each cut. The TS size and shape is altered during distinct periods of the solar cycle.}
    \label{fig:fig2.5}
\end{figure}

 Meridional slices in Figure \ref{fig:fig2.5} show the heliosphere during peak solar minimum (top row, 2008) and maximum (bottom row, 2015) of solar cycle 24. The innermost bold black line marks the TS, and the outermost line marks the HP. The supersonic solar wind plasma expresses typical latitudinal profiles, such as bimodal, fast and slow solar wind speeds accompanied by decreased and increased plasma density, respectively,  and large polar coronal hole regions (low B regions), during solar minimum. Alternatively, during solar maximum, the supersonic solar wind is replaced by more dense plasma in the high latitude regions with larger magnetic field strength. The plasma speed during solar maximum demonstrates the asymmetric nature of the solar wind out to the termination shock (\ref{fig:fig2.5}d), where faster solar wind can persist in the southern hemisphere. The slices also show how the termination shock expands and contracts, its shape evolving from solar minimum to maximum. We discuss the reaction of the TS and HP to the solar cycle in more detail in Section \ref{sec:3.4}.

\section{Results} \label{sec:3}

\begin{deluxetable*}{ccccc}
\tabletypesize{\footnotesize}
\tablewidth{\textwidth}
\tablecaption{Heliospheric Boundary Distances}
\tablehead{
\nocolhead{} &
\colhead{Steady-State} &
\colhead{Steady-State} &
\colhead{Time-Dependent} &
\colhead{Observations}  \\
\nocolhead{} &
\colhead{Single-Ion} &
\colhead{Multi-Ion} &
\colhead{Single-Ion} &
\nocolhead{}  \\
\nocolhead{} & \colhead{[au]} & \colhead{[au]} & \colhead{[au]}
& \colhead{[au]} } 
\startdata
TS (V1) & 87 $\pm$ 2 & 93 $\pm$ 1 & 81 $\pm$ 2 $\bm{^{2001.3}}$ & 94 $\bm{^{2005.0}}$\\
HP (V1) & 164 $\pm$ 2 & 149 $\pm$ 1 & 121 $\pm$ 2  $\bm{^{2012.4}}$& 122 $\bm{^{2012.7}}$ \\
TS (V2) & 81 $\pm$ 2 & 91 $\pm$ 1 & 76 $\pm$ 2 $\bm{^{2005.2}}$ & 84 $\bm{^{2007.7}}$  \\
HP (V2) & 143 $\pm$ 2 & 145 $\pm$ 1 & 113 $\pm$ 2 $\bm{^{2017.1}}$ & 119 $\bm{^{2018.9}}$\\
\enddata
\tablecomments{Values for single-ion and multi-ion are taken from the models of \citealt{chika2025} and \citealt{bair2025}, respectively. Bold superscripts represent the years of termination shock (TS) and heliopause (HP) crossings in the time-dependent solution and Voyager observations. Errors on modeled distances are calculated as twice the local grid cell size.}
\label{tab:tab1}
\end{deluxetable*}

\subsection{Plasma Flows and Magnetic Fields in the Heliosheath: Model Comparisons to Voyager Observations} \label{sec:3.1}

Models struggle to reproduce the size of the heliosphere, and locations of the heliospheric boundaries depend on the plasma flow and energy transport in the heliosphere \citep{Opher2023}. Including a time-dependent solar wind significantly improves the agreement of the model's heliospheric boundaries with observations. We find a V1 heliopause crossing of 121 au, in May 2012, that coincides with observations within the margin of error (Table \ref{tab:tab1}). For V2 the model predicts the heliopause crossing in February 2017, underestimating the heliopause distance by $\sim$6 au, which the analogous steady-state models overestimate by more than 24 au (Table \ref{tab:tab1}). The time-dependent model termination shock distances at V1 and V2 are both underestimated, because we do not separate pickup ions from the thermal solar wind as in the multi-ion model of \citealt{bair2025}. The heliosheath thickness can be further improved with a time-dependent, multi-ion model, and we plan to investigate the effects of such model in the future. We also note that the termination shock and heliopause locations are dependent on the strength and direction of $\textbf{B}_{ISM}$ which are still being constrained (e.g. \citealt{opher2009, provornikova2014}). 

\begin{figure}[ht!]
    \centering
    \includegraphics[width=1\linewidth]{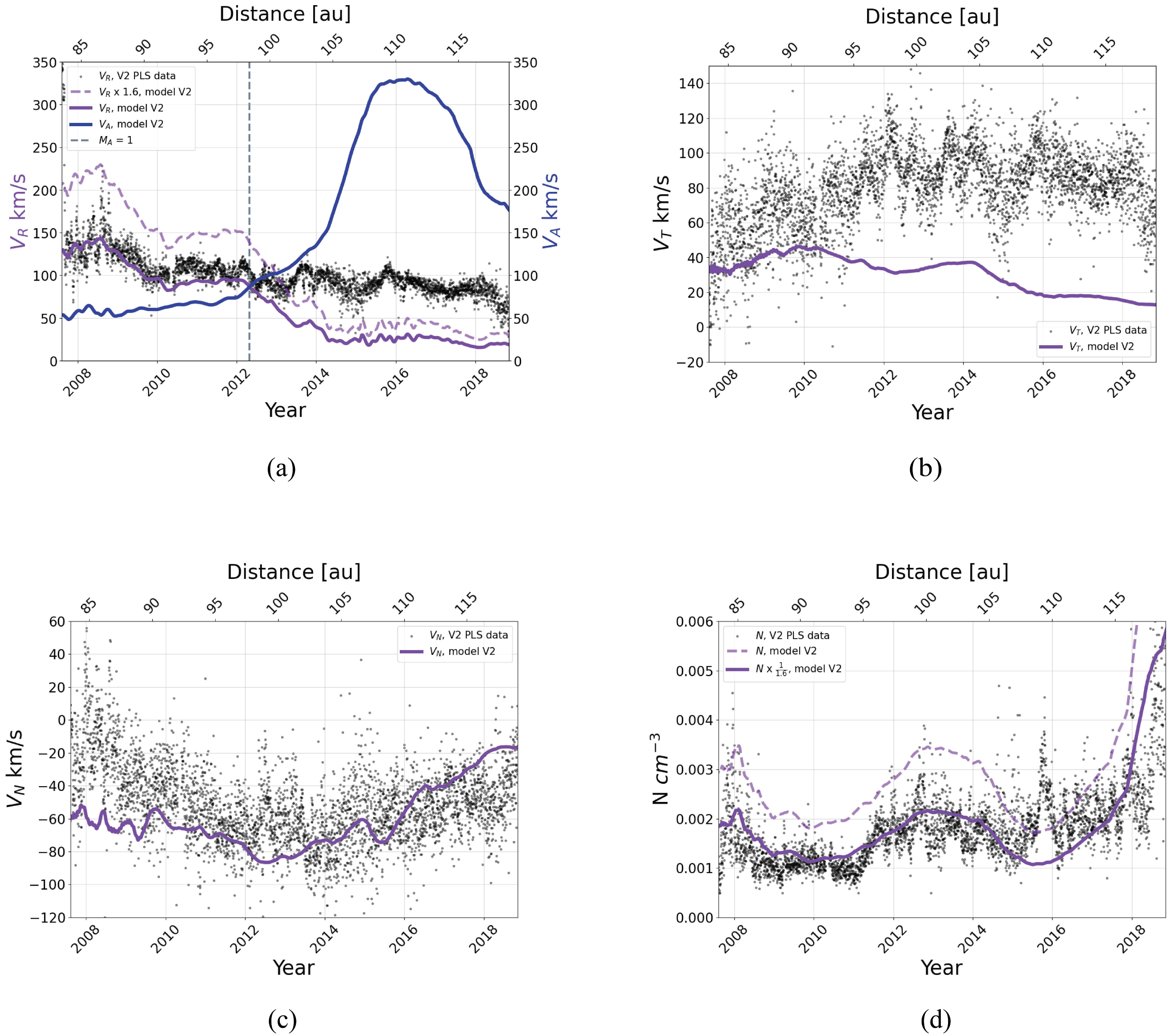}
    \caption{Plasma (a) radial velocity and Alfvén speed, (b) tangential velocity, (c) normal velocity, and (d) number density are shown in comparison to V2 observations in black points. The model results are aligned to the heliopause, as observed by V2 and solid lines represent the plasma output that match V2 measurements immediately downstream of the termination shock. We reduce the density by a factor 1.6 to match the data, showing the unmodified density profile in the dashed line of (d). We also show the radial velocity increased by a factor of 1.6 in the dashed line of (a) for completeness. The vertical grey line in (a) marks where the simulated flows become sub-Alfvénic at Alfvén Mach number, $M_A$ = 1.}
    \label{fig:fig3}
\end{figure}

Figure \ref{fig:fig3} shows comparisons of the time-dependent model's velocity and density profiles to observations from V2 in the heliosheath. We shift the model results in time to align with observed heliopause locations and evaluate the solid lines in Figure \ref{fig:fig3} which best match the V2 plasma measurements immediately downstream of the termination shock. In plasma density we reduce the results by a factor of 1.6 to account for the strong termination shock predicted by the single-ion treatment (Figure \ref{fig:fig3}d, solid line). Without separated pickup ions, the plasma temperature and magnetosonic speed are lower in the supersonic solar wind, resulting in a strong shock. This effect has been resolved in models with a multi-ion treatment of thermal solar wind and pickup ions (Opher et al. 2020) and we will combine this treatment with time-dependent effects in the future. Although the model predicts a stronger termination shock, the jump in plasma speed occurs when the solar wind is faster than observed because the termination shock location in the single-ion treatment is closer to the Sun than observations show. Because of this we do not need to scale the radial velocity downstream of the termination shock to match V2 measurements, but we show the radial velocity increased by factor of 1.6, to reciprocate the scaling applied to the plasma density (Figure \ref{fig:fig3}d, dashed line).

The model reproduces short, fluctuating peaks occurring every 0.5 to 1 year in data observed by V2 in the radial velocity ($V_{R}$), tangential velocity ($V_{T}$), and normal velocity ($V_{N}$) components in the heliopause-aligned comparison (Figure \ref{fig:fig3}a,b,c, solid lines). Solar-cycle plasma speed variations capture the smooth step-like decrease in $V_{R}$ during 2009 that appears in the data. However, $V_{R}$ continues to decrease towards the heliopause, eventually falling a factor of 3 below the observed radial velocity profile. Our results are consistent with the findings of other time-dependent studies that showed the same decrease in $V_{R}$ toward the heliopause \citep{provornikova2014,izmo&alex2020}, implying that solar cycle variations in the speed are not the source of sustained radial velocities measured at V2 near the heliopause. $V_{T}$ and $V_{N}$ are very sensitive to the heliosphere shape \citep{opher2012,provornikova2014} and our model produces a heliosphere that is more blunt in the normal direction than in the tangential direction, with $V_{N}$ speeds larger than $V_{T}$ by 36 km s$^{-1}$ throughout most of the heliosheath. However, V2 observations show the opposite where $V_{T}$ remains larger than $V_{N}$. 

The $V_{N}$ profile is mostly consistent with the V2 data; the remaining inconsistencies in $V_{R}$ and $V_{T}$ could be partly attributed to the single-ion description of the plasma in the simulation. Pickup ions treated separately shrink the heliosheath which can affect the flows \citep{opher2020,opher2025}. However, the effects of pickup ions alone are not sufficient to resolve the differences in plasma speeds. We also eliminate mass loading as a possible contributor to the disagreement between the simulated plasma flows and observations because the progression of the scaled plasma density in the heliosheath reproduces the V2 density profile (Figure \ref{fig:fig3}d, solid line). The model captures the decrease in density after 2012 and increase before the heliopause after 2016; we find no discrepant features in density that could be associated with the underestimated flows in $V_{R}$ and $V_{T}$. 

\begin{figure}[ht!]
    \centering
    \includegraphics[width=1\linewidth]{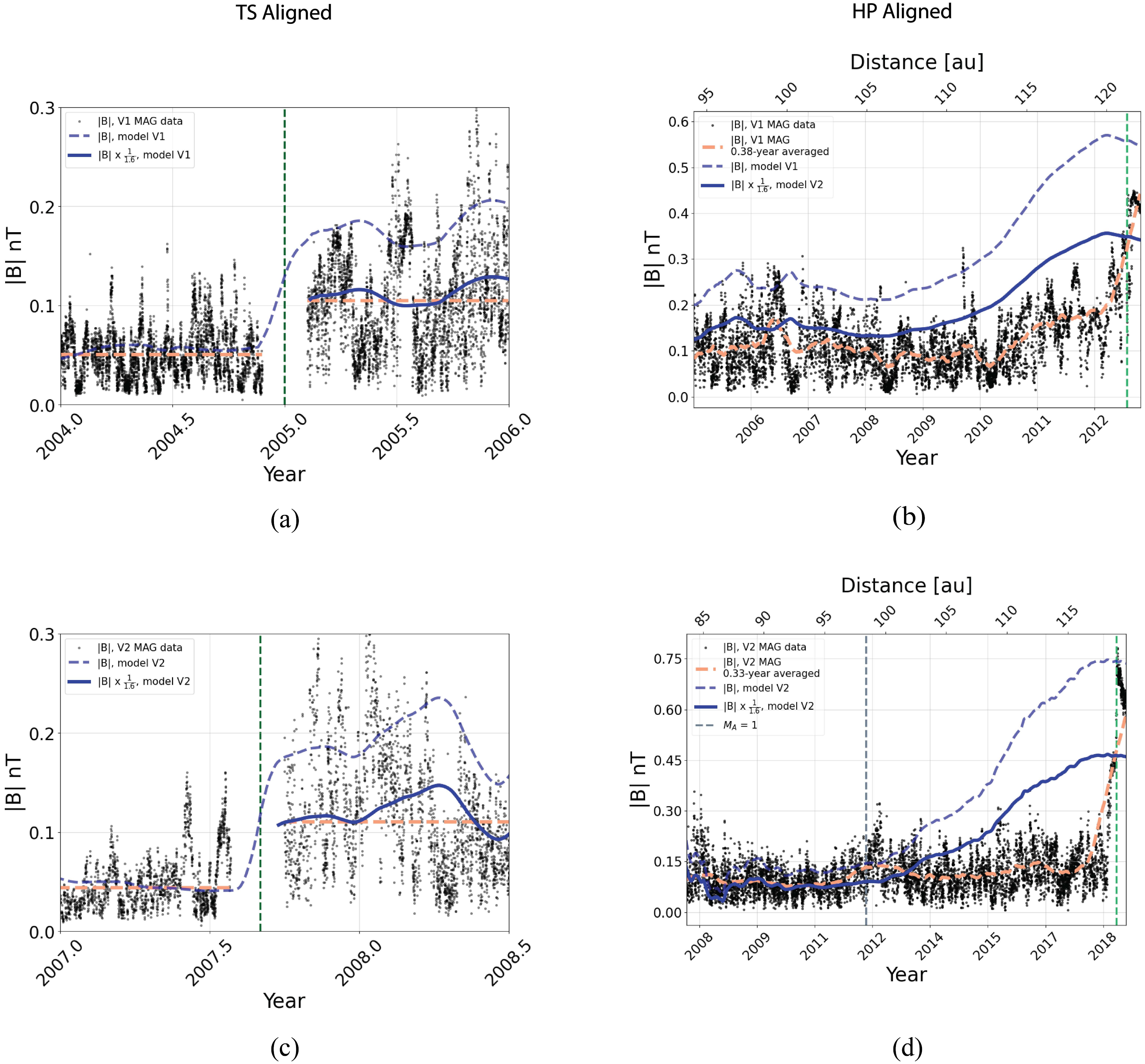}
    \caption{Model results of the magnetic field profile in comparison to the Voyager observations in black points. (a,c) We show how the scaling applied to the output at the termination shock; the magnetic field is reduced by a factor of 1.6 to match the Voyager measurements averaged over 1 year immediately downstream of the termination shock, shown by the orange dashed line. Solid and dashed lines blue lines represent the scaled and unmodified magnetic field profiles, respectively. (b,d) The magnetic field profiles in the heliosheath are aligned to the heliopause crossings observed by Voyager. The trends in the observations of underscored by the orange dashed line representing 0.38-year and 0.33 year averages of the V1 and V2 data, respectively. The model develops a pile-up region $\sim$15 au ahead of the heliopause. The grey, vertical dashed line along V2 represents the sub-Alfvénic transition from Figure \ref{fig:fig3}a.}
    \label{fig:fig4}
\end{figure}

The radial velocity is correlated to the Alfvén speed, which dramatically increases around 2012.3 (Figure \ref{fig:fig3}a, solid blue and dashed grey lines), coincident with the radial velocity deviating from the observations. Heliosheath data do not show a sub-Alfvénic region, indicating that magnetic energy in the model is in excess of realistic conditions. We compare trends in the time-dependent magnetic field profile to Voyager observations in the heliosheath in Figure \ref{fig:fig4}b and d, shifted in time to match the Voyager heliopause crossings. As with density, we reduce the model's downstream magnetic field strength by a factor of 1.6, to match the average magnetic field measured by Voyager behind the termination shock (Figure \ref{fig:fig4}, solid lines). The scaled magnetic field gives good agreement to the data, until $\sim$15 au prior to the heliopause crossing (Figure \ref{fig:fig4}b,d), beyond which the magnetic field strength gradually increases, overestimating observations by approximately a factor of 2 for V1 and up to a factor of 3 for V2. The plasma radial velocity decreases as a consequence of the elevated magnetic field strength, which corresponds to the location where flows along V2 become sub-Alfvénic in Figure \ref{fig:fig3}a. The plasma beta, the ratio of the thermal pressure to magnetic pressure, becomes $<$1 in the sub-Alfvénic region which is also found in steady state model's with separated pickup ions \citep{bair2025}. Magnetic fields can alter the plasma flow in a low beta plasma regime \citep{izmod2015}, and V1 and V2 measurements of the heliosheath have been characterized with a plasma beta that is always $>$1 \citep{dialynas2017a,dialynas2019}. However, within 1 au of the heliopause, energetic particle measurements from V2 showed that the plasma beta becomes $<$1 \citep{krimigis2019}. We find a low beta region much wider than V2 observations show due to the magnetic field that piles up and increases in front of the heliopause, despite the time-dependent variations applied to the magnetic field intensity. This may indicate that magnetic reconnection, active in the region before the heliopause, is missing from the model \citep{izmo&alex2020}. Simulated magnetic reconnection may be needed in the heliosheath (\citep{drake2010, opher2011} as a mechanism to diminish energy in the magnetic field. \citealt{korolkov2025} showed that modeling the large-scale effects of magnetic reconnection can better reproduce the Voyager's magnetic field observations. The role of magnetic reconnection on acceleration of particles, as well as the observable effects, will be explored in a future study. The consequences of removing magnetic field also needs to be investigated because \citealt{drake2015} showed that flows in the heliosheath can still be controlled by the magnetic field if the magnetic pressure is comparable to the dynamic pressure and magnetic tension is sufficiently strong.

\subsection{Solar Transients in the Heliosheath and Interstellar Medium: Plasma Pulses and Magnetosonic Waves} \label{sec:3.2}
\begin{figure}[ht!]
    \centering
    \includegraphics[width=1\linewidth]{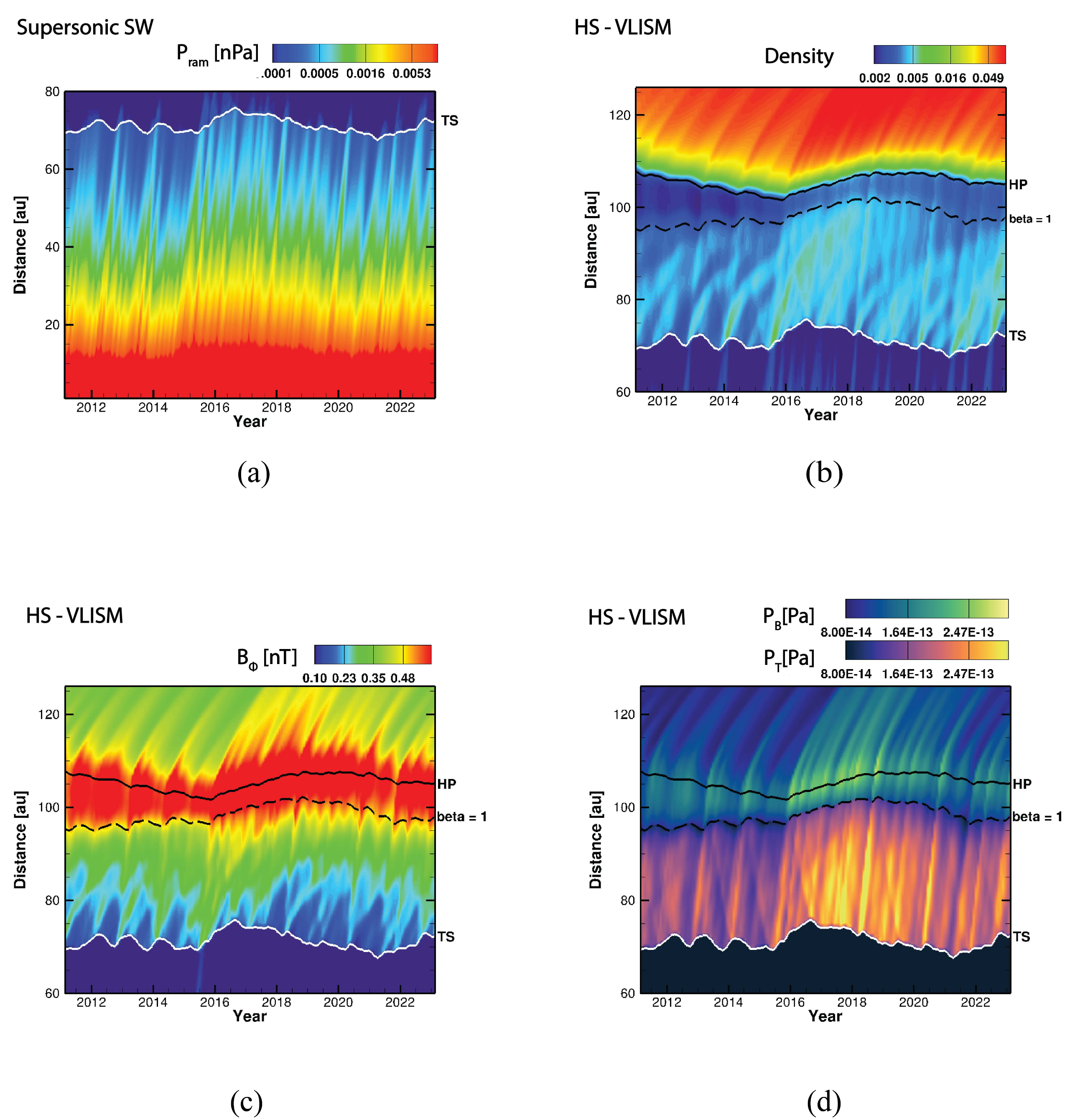}
    \caption{Upwind spacetime plots showing the (a) ram pressure in the supersonic solar wind (SW) and the (b) plasma number density, (c) azimuthal magnetic field, and (d) thermal and magnetic pressure in the heliosheath (HS) and very local interstellar medium (VLISM). Solid black and white lines show the termination shock (TS) and heliopause (HP), respectively. The magnetic pile-up region, roughly outlined by the dashed black line at $\beta = 1$, forms around 100 au and its spatial extent varies with time. Plasma pulses in the magnetic field, density, and pressure emerge downstream of the termination shock. These structures are associated with magnetosonic waves that propagate through the heliosheath. (d) The layered spacetime plot, showing thermal pressure in purple-orange color and magnetic pressure in blue-green color, illustrates how pressure waves in the heliosheath are transmitted across the heliopause.}
    \label{fig:fig5}
\end{figure}

Plasma pulses in the heliosheath were shown with spacetime plots first by \citealt{washimi2011} during the period of 2002 to 2009. We expand on those results showing the pulses in the heliosheath during the most recent solar cycle (solar cycle 24) from 2011 to 2023. In Figure \ref{fig:fig5}, the bold white line around 70 au marks the location of the termination shock, and the bold black line around 105 au marks the heliopause. The magnetic pile-up region is represented by the dashed, $\beta = 1$ boundary, across which the plasma becomes low beta. Pulses in the heliosheath are driven by dynamic or ram pressure pulses in the supersonic solar wind, starting from the inner boundary of the model at 1 au (Figure \ref{fig:fig5}a) which evolve with distance. Because we capture a full solar cycle, we see pulses occur every 0.3 to 1 year in the heliosheath, and their recurrence decreases following solar minimum (2011-2015) and increases following solar maximum (2015-2020). Our results are consistent with V2 observations of plasma pulses, with lengths of 3-6 months, that result in a factor of 2 increase in density and up to a factor 3 increase in total pressure in the heliosheath \citep{richardson2017}. 

Density in the heliosheath (Figure \ref{fig:fig5}b) is sensitive to slower moving pulses, their trajectories possessing a more shallow slope in the spacetime domain. These pulses propagate to the magnetic pile-up region, afterwards diminishing and flattening before reaching the heliopause. Associated pulses are also transported through the magnetic field, shown by the azimuthal component in Figure \ref{fig:fig5}c, with a similar slope as those present in density, although the pile-up region slightly obscures the structure further in the heliosheath. Pulse structures in the magnetic field also advect out into the local interstellar medium. We connect these oscillating structures in the gas and magnetic field components with a layered spacetime plot in Figure \ref{fig:fig5}d. The layered spacetime plot shows the thermal pressure immediately downstream the termination shock in purple-orange color, followed by the magnetic pressure in blue-green at the $\beta = 1$ boundary. Thermal and magnetic pressures are sensitive to a faster moving pulse than those in density, with higher recurrence than its slower moving counterpart. As the solar wind approaches $\beta = 1$, pulses in the thermal pressure are extinguished due to the increase in the magnetic field strength, but they continue propagating unaffected through the magnetic field as they transfer seamlessly into magnetic pressure. These faster moving pulses propagate backwards towards the terminations shock, as well as persist across the heliopause, into the local interstellar medium.

\begin{figure}[ht!]
    \centering
    \includegraphics[width=1\linewidth]{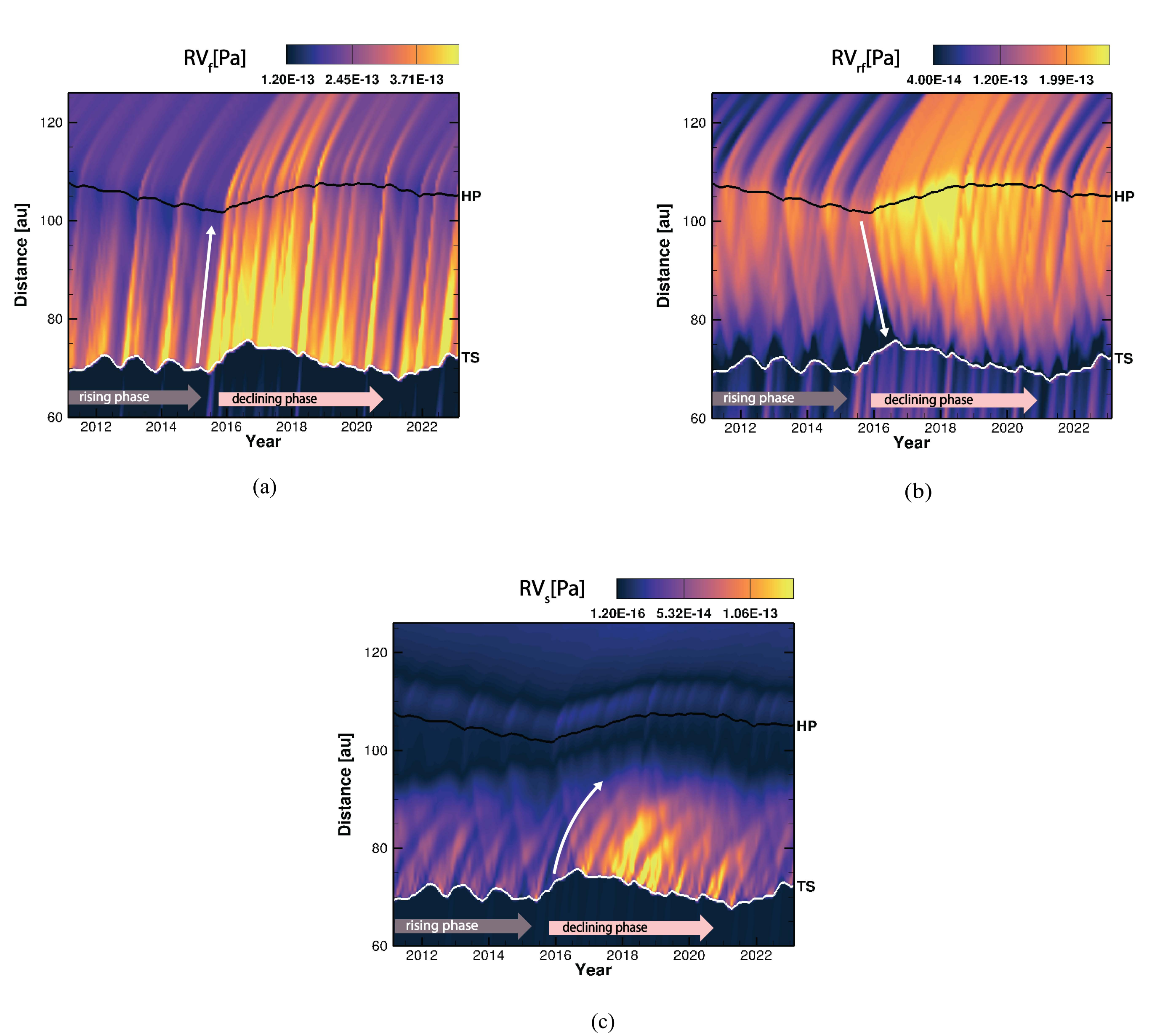}
    \caption{Linear Riemann variables (RV) characterizing magnetosonic wave propagation in the heliosheath. (a) $RV_{f}$, (b) $RV_{rf}$, and (c) $RV_{s}$ represent the trajectories of outgoing fast mode, reflected fast mode, and slow modes waves, respectively. Magnetosonic waves are more prevalent in the declining phase of solar activity, which follows solar maximum. These waves explain the pulse features in Figure \ref{fig:fig5}.}
    \label{fig:fig6}
\end{figure}

Works that analyzed these plasma pulses identified them as magnetosonic pulses or entropy waves based on their speed \citep{storyzank1997, washimi2011}. Additonally, \citealt{zank2017} demonstrated that compressible magnetic fluctuations in the local interstellar medium, observed by V1, can be explained by heliosheath fast and slow mode waves incident on the heliopause. We distinguish the plasma pulses explicitly as fast and slow magnetosonic waves characterized by linear Riemann variables, or RVs (see derivation in the Appendix)for the first time. Spacetime plots of the RVs for fast mode ($RV_f$), reflected fast mode ($RV_{rf}$), and slow mode ($RV_s$) waves are shown Figure \ref{fig:fig6}. Interplanetary transients in the super sonic solar wind, shown in Figure \ref{fig:fig5}a, perturb the termination shock, producing oscillations that generate fast and slow mode waves in the heliosheath (described previously by \citealt{washimi2011}). Wave paths are drawn out by $RV_f$ and $RV_{rf}$ in the heliosheath align with the steep, thermal and magnetic pressure pulse structures; we identify those pressure pulses, specifically, as manifestations of compressional fast mode waves (Figure \ref{fig:fig6}a). Moreover, we attribute density pulses in the heliosheath, to slow mode waves shown by $RV_s$ in Figure \ref{fig:fig6}c, as both structures have matching trajectories. Like the plasma pulses, both wave modes have clear solar-cycle dependence: less frequent (every 1 year) in the rising phase of solar activity, and more frequent (every 0.3 year) in the declining phase. Fast modes are generally more prevalent in the heliosheath, especially in the declining phase. We find slow modes have a propagation speed around 93 km s$^{-1}$, consistent with the slow magnetosonic speed, or solar wind speed, in this region. The solar wind radial velocity decreases as it approaches the heliopause, and similarly slow modes decelerate and damp with increasing distance. Because of this, these modes do not survive long in the heliosheath and have no interaction with the heliopause. Alternatively, fast modes travel around 360 km s$^{-1}$ in the heliosheath, consistent with the local fast magnetosonic speed, and propagate much further than slow modes. They are able to collide with the heliopause, causing the boundary to oscillate. Fast modes also dominate because they have a counter-propagating component, represented by $RV_{rf}$ in Figure \ref{fig:fig6}b. Fast modes reflected at the heliopause are directed back towards the termination shock with speeds around -197 km s$^{-1}$. As a result, the termination shock is not only influenced by upstream interplanetary structures, but also reflected waves that force the boundary inwards. Fast modes are long-lived; we show that they also persist into the interstellar medium, transmitting and refracting strongly across the heliopause (Figure \ref{fig:fig6}a). Our results match the description of fast modes wave progression from the heliosheath, to the interestllar medium, described in \citealt{zank2017}.

\subsection{The Origin of Pressure Fronts Observed in the Interstellar Medium} \label{sec:3.3}

V1 observed pressure fronts in the local interstellar medium in 2017 and 2020 when the magnetic field strength jumped by a factor 1.19 and 1.35, respectively \citep{burlaga2021}. These events have been referred to as pressure front 1 and 2, or pf1 and pf2. The source of both pressure fronts has been speculated on since their observation, especially pf2 which was much stronger than any preceding jump feature. We utilize the fast mode wave characteristic, $RV_{f}$, to trace back pressure fronts to 1 au.

\begin{figure}[ht!]
    \centering
    \includegraphics[width=1\linewidth]{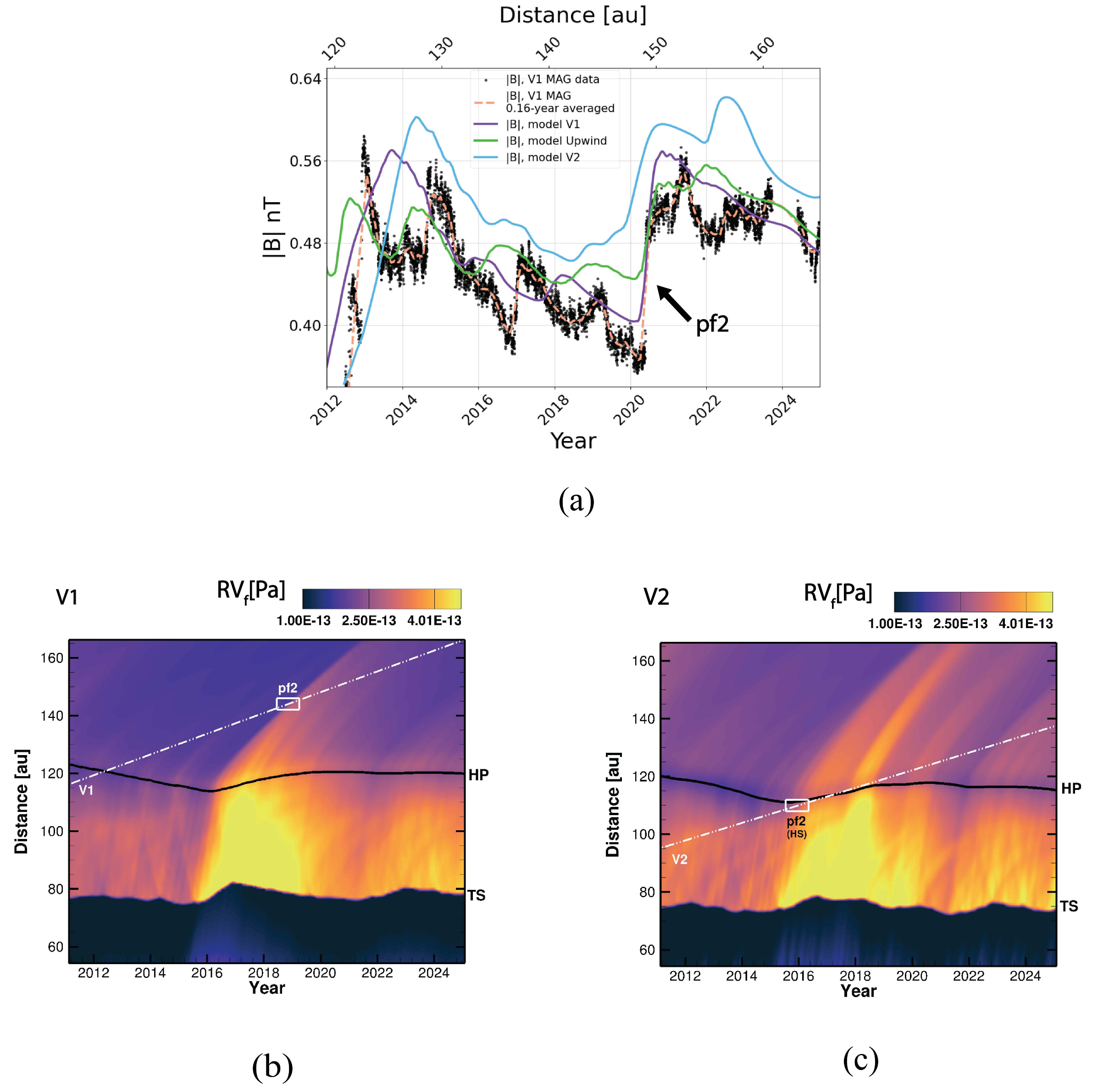}
    \caption{(a) Simulated magnetic field strength in the interstellar medium along the V1 trajectory (purple line), and the V1 trajectory project in the V2 (blue line) and upwind (green line) directions in comparison to V1 observations (black points). The orange dashed line shows the data in 2-month averaged windows. The large jump in magnetic field around 2020 is pf2, and we locate the jump in each spacecraft direction. Spacetime plots of $RV_{f}$ show the pf2 location in the (b) V1 direction and (c) V2 direction determined by the fast mode wave characteristics. The Voyager spacecraft trajectories in (b) and (c) represented by the dashed white lines intersect pf2 in late 2018 along V1, and late 2015 in along V2.}
    \label{fig:fig6.1}
\end{figure}

\begin{figure}[ht!]
    \centering
    \includegraphics[width=1\linewidth]{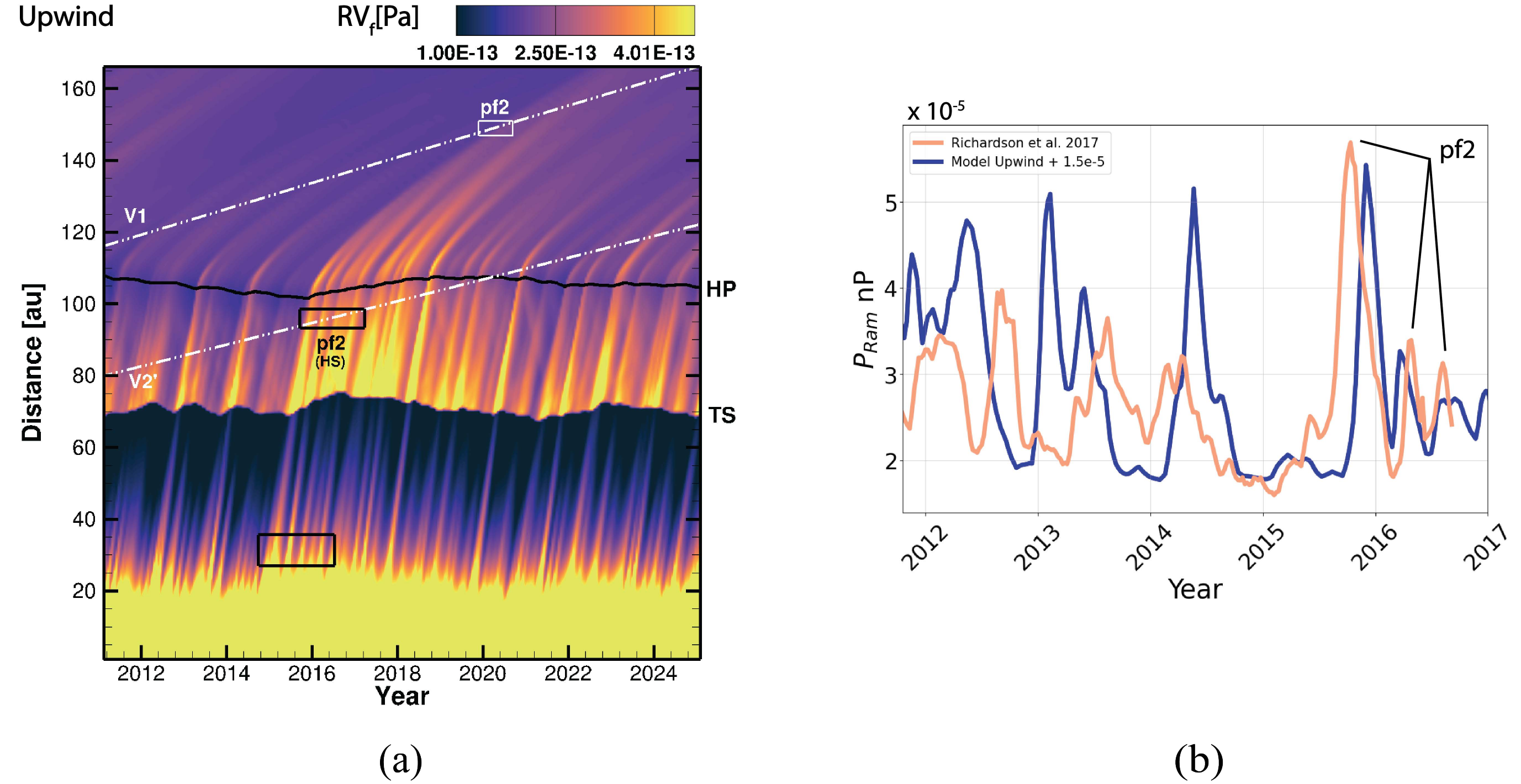}
    \caption{Tracing back the pf2 feature identified in Figure \ref{fig:fig6.1}b. (a) Five pressure pulses in the supersonic solar wind from 2014.7 - 2015.9, outlined in the black box around 30 au, are the origin of pf2. This group of pulses reach the termination shock around 2015.6 - 2016.6,  propagate as fast mode waves in the heliosheath, and reach the heliopause around 2016.1-2017.1. Once crossing the heliopause, pulses merge in the interstellar medium before reaching the actual pf2 location. (b) Sampling the ram pressure along the V2-like trajectory upwind, labeled V2' (a), gives good agreement to V2 observations of pressure pulses in the heliosheath. These pulses eventually merge to become pf2.}
    \label{fig:fig6.2}
\end{figure}

In comparison to magnetic field observations in the local interstellar medium, time-dependent variations can produce jump features similar to those seen by V1 (Figure \ref{fig:fig6.1}a). In particular, we identify pf2 in the simulated magnetic field, the largest jump near 2020, along the V1 trajectory (purple line), and the V1 trajectory projected in the V2 (blue line) and upwind (green line) directions. The magnetic field profiles in the V1 and V2 directions are shifted forward by 1.5 years and 2.4 years, respectively, to match the timing of the pf2 jump in the data. We locate the pf2 event in spacetime plots of Figures \ref{fig:fig6.1}b and c, highlighting its location along V1 in the interstellar medium, and along V2 in the heliosheath. In the V1 direction, pf2 is coincident with a large fast mode wave that is transmitted from the heliosheath, produced during a period of high solar activity in solar cycle 24. In the V2 direction, we see that the spacecraft encountered this wave while it was transiting the heliosheath. Because of limited temporal resolution at V1 and V2 latitudes in the model (shown in Figure \ref{fig:fig2}b) we cannot fully distinguish the structure of pf2 in the heliosheath or interstellar medium in the V1 and V2 directions; we can only establish the time of pf2's incidence at each spacecraft. However, pf2 and other pressure fronts in general are, in general, heliosphere-sized structures, given that we see related jumps in the magnetic field (Figure \ref{fig:fig6.1}a) in several directions. Upwind, where the model has high temporal resolution, we show that pf2 is the result of a series of fast mode waves, in a process that propagates from 1 au to the local interstellar medium (Figure \ref{fig:fig6.2}a). Near 1 au, pf2 originates as five discrete pressure pulses from 2014.7 to 2015.9, each pulse emerging $\sim$3 months apart. These likely form global merged interaction regions (GMIRs), which become more frequent during periods of high solar activity \citep{richardson2003}. The pulses take a year to reach the termination shock, arriving at the boundary from 2015.6 to 2016.6, still as five separated structures. Once crossing the termination shock, they are amplified and propagate as fast mode waves in the heliosheath as discussed in Section \ref{sec:3.2}. The ram pressure along a V2-like trajectory upwind (labeled V2' in Figure \ref{fig:fig6.2}a) suggests that V2 observed this group of waves ( which eventually form pf2) in the heliosheath around 2016, in the form of pressure pulses; their timing agrees well with V2 observations from \citealt{richardson2017} (Figure \ref{fig:fig6.2}b). Waves cross the heliosheath in half a year, reaching the heliopause by 2016.1 to 2017.1, and refracting across the boundary. 
The speed of fast mode waves decreases by $\sim$288 km s$^{-1}$ upon entering the interstellar medium; in this region we see the propagation speed of succeeding waves increases. Because of this, following waves overtake and merge with one another, increasing in amplitude by the time they intersect V1's location in the interstellar medium. The compressional effect of this composite wave is a large enhancement in the magnetic field, which may correspond to pf2. Our results show that merging of waves in the interstellar medium is a common process, powered by solar cycle variations. Other pressure fronts in the interstellar medium arise by the same process, but pf2 was exceptionally large as it was the product of five (or more) coalesced fast modes waves, whereas weaker pressure fronts, like pf1 1 in 2017, result from only two or three waves. \citealt{burlaga2021} made a similar explanation that pf2 resulted from two pressure fronts overtaking each other. Previous time-dependent model results from \citealt{pogorelov2021} and \citealt{zirnstein2024} also showed that series of shocks or pressure pulses will merge in the interstellar medium by an analysis of one dimensional magnetic field profiles. The data-driven model of \citealt{fraternale2026} also reproduces the pf2 jump, and suggests that the event is the result of two merged shocks or compressions that traveled through the heliosphere. Based on the fast mode wave characteristics, our results suggest at least five compression features merge to produce pf2, and the origin of those features at 1 au from 2014 to 2016 complement the findings of \citealt{fraternale2026}.

\subsection{Behavior of Heliospheric Boundaries and Implications for New Horizons' Termination Shock Crossing} \label{sec:3.4}

\begin{figure}[ht!]
    \centering
    \includegraphics[width=1\linewidth]{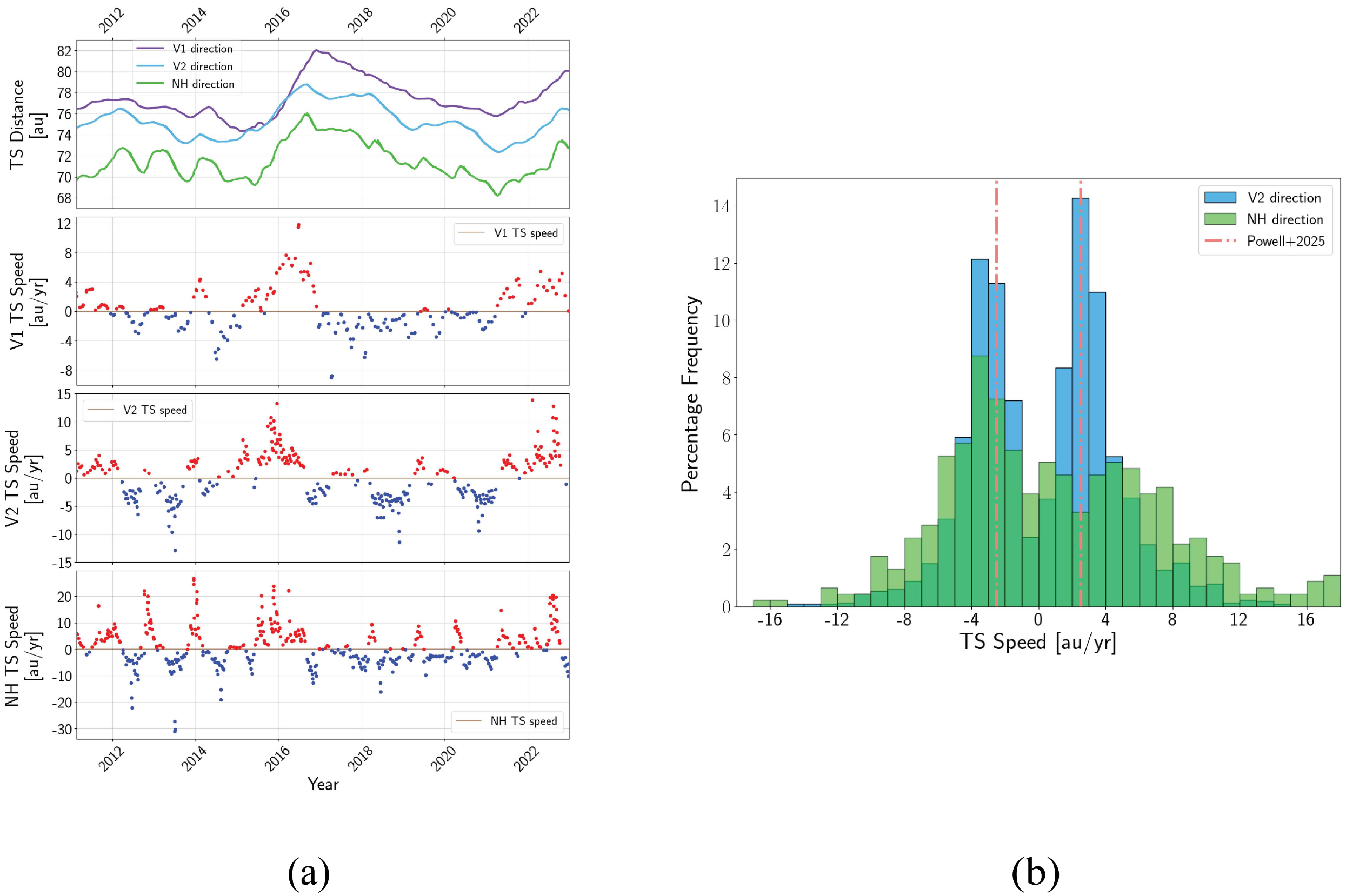}
    \caption{(a) Time-dependent model solutions for the termination shock distance and speed from 2011 to 2023. The top panel shows the shock location in the V1 (purple line), V2 (blue line), and NH (green line) directions. The speed of the shock in both directions is displayed in the second and third panels, with outward radial speeds represented by red points, and inward speeds by blue points. Each panel highlights the large variability in the termination shock as a consequence of solar activity. (b) The distribution of termination shock speeds along V2 and NH agree with speed estimates made by \citealt{powell2025} (orange dashed line).}
    \label{fig:fig7}
\end{figure}

The termination shock and heliopause boundaries fluctuate on long and short time scales. Long scale variations refer to 11-year, solar cycle periodic changes, whereas short scale variations are related to solar transients. In the NH direction, our model output in the top panel of Figure \ref{fig:fig7}a shows that from 2011 to 2023 the termination shock has an average heliocentric distance of 71.5 au, with 7.8 au excursions from minimum to maximum distance. In the same time range, along the V1 and V2 directions the termination shock has an average distance of 77.5 au with 7.7 au total excursion, and 75.3 au with 6.4 au total excursion, respectively. These values are indicative of long time scale variations: the termination shock reaches its maximum distance in late 2016 following solar maximum in 2015, and attains its minimum distance in 2021 after solar minimum in 2020 for each direction. The termination shock relaxes gradually in the declining phases of solar activity (2016-2021) as the dynamic pressure in the supersonic solar wind diminishes and increased thermal pressure contained in the heliosheath drives the shock inward. Generally, the termination shock spends more time receding than expanding. For instance, the termination shock begins a sharp jump outward in mid-2015, which lasts for 1 year, then moves slowly inward over a 5 years period. This behavior is consistent with observations shown by \citealt{sokol2021} of the solar wind dynamic pressure at 1 au. Dynamic pressure peaks quickly at solar maximum, then gradually falls throughout the declining and rising phase of the next solar cycle. The amplitude and fluctuating features in our model's termination shock motion bears similarities to the time-dependent model of \citealt{izmo&alex2020} from 2011 to 2018. The variation, from minimum to maximum distances, is also in agreement to the total termination shock distance variation reported by other time-dependent models (e.g. \citealt{zankmuller2003}, \citealt{izmodenov2008}) of $\sim$7 au in the upwind direction. Time-dependent analysis from \citealt{washimi2011} reported larger variations of 12-13 au in the termination shock from 2001-2008. However, during that period, the relative increase in dynamic pressure in the heliosphere was considerably greater than the dynamic pressure in solar cycle 24 (2008-2019). Current dynamic pressure trends in solar cycle 25 are most similar to solar cycle 24 in amplitude (OMNI near-Earth solar
data\footnote{\url{http://omniweb.gsfc.nasa.gov/}}), and this is the time range we include in Figure \ref{fig:fig7}a. However, forecasts for solar wind dynamic pressure variations in time by Gasser et al. 2026 (in press) indicate that the solar wind dynamic pressure will stay at elevated values for the remainder of the solar cycle 25.

Short scale variations in the termination shock location are carried out by continual perturbations in the plasma due to pressure pulses in the heliosphere. When pulses collide with the termination shock, either from the supersonic solar wind or as a reflected fast mode wave in the heliosheath, the shock responds with rapid outward or inward motion. The second, third, and fourth panels of Figure \ref{fig:fig7}a show the instantaneous termination shock speed for each spacecraft direction. Red model points represent positive radial speeds (outward) and blue points represent negative radial speeds (inward). The termination shock is highly variable, with an average radial speed of 6.05 au yr$^{-1}$ $\pm$ 5.37 au yr$^{-1}$ in the NH direction from 2011 to 2023. Speeds in the V1 and V2 direction are slower and less variable in our model, with average speeds of 2.16 au yr$^{-1}$ $\pm$ 2.04 au and yr$^{-1}$ 3.59 au yr$^{-1}$ $\pm$ 2.32 au yr$^{-1}$, respectively. Lower speeds in the Voyager directions may partly be a consequence of our model's low temporal resolution, but we do not eliminate the possibility that the termination shock motion may vary less at higher latitudes in actuality. Observations have shown that fast solar wind at higher latitudes tends to be more homogenous over time \citep{mccomas2008}, than solar wind in the ecliptic plane where NH resides.

In the fourth panel of Figure \ref{fig:fig7}a, along NH's trajectory the model shows that in rising phases of the solar cycle (2011 to mid 2015, following solar minimum) the termination shock motion has higher speeds than during the declining phases (mid 2015 to 2021, following solar maximum). The termination shock travels outward up to 14 au yr$^{-1}$ in the rising phase of the solar cycle, coincident with the arrival of three prominent plasma pulses, occurring about one year apart. Averaged termination shock speeds along the NH trajectory in Table \ref{tab:tab2} also demonstrate that speeds in the rising phases are almost double of those in the declining phase. Pressure pulses take roughly 1.6 years to transit out and back in the heliosheath. In that time, the shock has about half a year to exhibit 2 au variations in response to outgoing and reflected pulses. As a result, the termination shock motion in the rising phase is sinusoidal-like. Conversely, pressure pulses in the declining phase, just after solar maximum, are abundant, emerging and reflecting in the heliosheath just months apart (Figure \ref{fig:fig6}a,b). While increased solar activity produces more variability in the plasma environment itself, high recurrence of outgoing and reflected pulses create a stabilizing effect on the termination shock. Because of this, the termination shock exhibits short variations of only 1 au or smaller in the declining phase, and long scale behavior primarily dictates the shock behavior in this phase. Unlike in the NH direction, termination shock speeds between the rising and declining phases are generally consistent in the Voyager directions because the short scale variations are not resolved at those latitudes (Table \ref{tab:tab2}). However, the termination shocks speeds tend to be greater in the V2 direction compared to V1, as seen in Table \ref{tab:tab2} and second and third panel of Figure \ref{fig:fig7}a, suggesting that the southern heliosphere may have experienced more dynamic solar wind conditions during this particular period of solar cycle 24.

\begin{deluxetable*}{ccc}
\tabletypesize{\footnotesize}
\tablewidth{\linewidth}
\tablecaption{Average TS speeds, along NH and V2 directions, during rising and declining phases of the solar cycle.}
\tablehead{\\
\colhead{Speed} &
\colhead{\hspace{.75cm}Rising Phase}\hspace{.5cm} &
\colhead{\hspace{.33cm}Declining Phase}\hspace{.33cm} \\
\nocolhead{} & \colhead{[au/yr]} & \colhead{[au/yr]}} 
\startdata
\vspace{0.2cm}
Inward (NH) &  -6.21 $\pm$ 6.29 & -4.20 $\pm$ 2.84 \\
\vspace{0.2cm}
Inward (V1) & -1.70 $\pm$ 1.53 & -1.91 $\pm$ 1.80 \\
\vspace{0.2cm}
Inward (V2) & -3.38 $\pm$ 1.93 & -3.79 $\pm$ 2.42  \\
\vspace{0.2cm}
Outward (NH) & 7.52 $\pm$ 6.56 & 4.74 $\pm$ 3.01  \\
\vspace{0.2cm}
Outward (V1) & 1.45 $\pm$ 1.23 & 2.83 $\pm$ 2.38 \\
\vspace{0.2cm}
Outward (V2) & 2.28 $\pm$ 1.24 & 1.72 $\pm$ 0.92 \\
\enddata
\tablecomments{Averages and standard deviations are calculated in the time range of 2011 to 2015 for the rising phase, and 2017 to 2021 for the declining phase. The model points used for these calculations are the same as shown in the bottom two panels of Figure \ref{fig:fig7}a}
\label{tab:tab2}
\end{deluxetable*}

Figure \ref{fig:fig7}b shows the distribution of termination shock speeds in the NH (blue bins) and V2 (pink bins) directions. The distributions are bimodal, centered around 0 au yr$^{-1}$. The major mode in the distribution along NH peaks at negative radial speeds, around -4 au yr$^{-1}$, reiterating that the termination shock spends more time deflating than expanding. From Table \ref{tab:tab2}, we know that this inward speed of -4 au yr$^{-1}$ is primarily occurring during the declining phase of the termination shock motion. \citealt{powell2025}, using V2 observations of termination shock particles (TSPs) and modeling the magnetic connectivity of V2 to the shock in steady-state, inferred a speed of $\pm$2.5 au yr$^{-1}$. The estimates from this works are marked with dashed vertical lines in Figure \ref{fig:fig7}b and they agree with the possible speeds produced by our model, especially at the peaks in the distribution along V2. Values from \citealt{powell2025} correspond to even longer time scales as TSP observations at V2 were made months before the shock crossing. This comparison shows that a dynamical simulation, which captures the short and long scale changes in the solar wind, reveals much faster termination shock speeds than what can be inferred by steady-state models. 

\begin{figure}[ht!]
    \centering
    \includegraphics[width=1\linewidth]{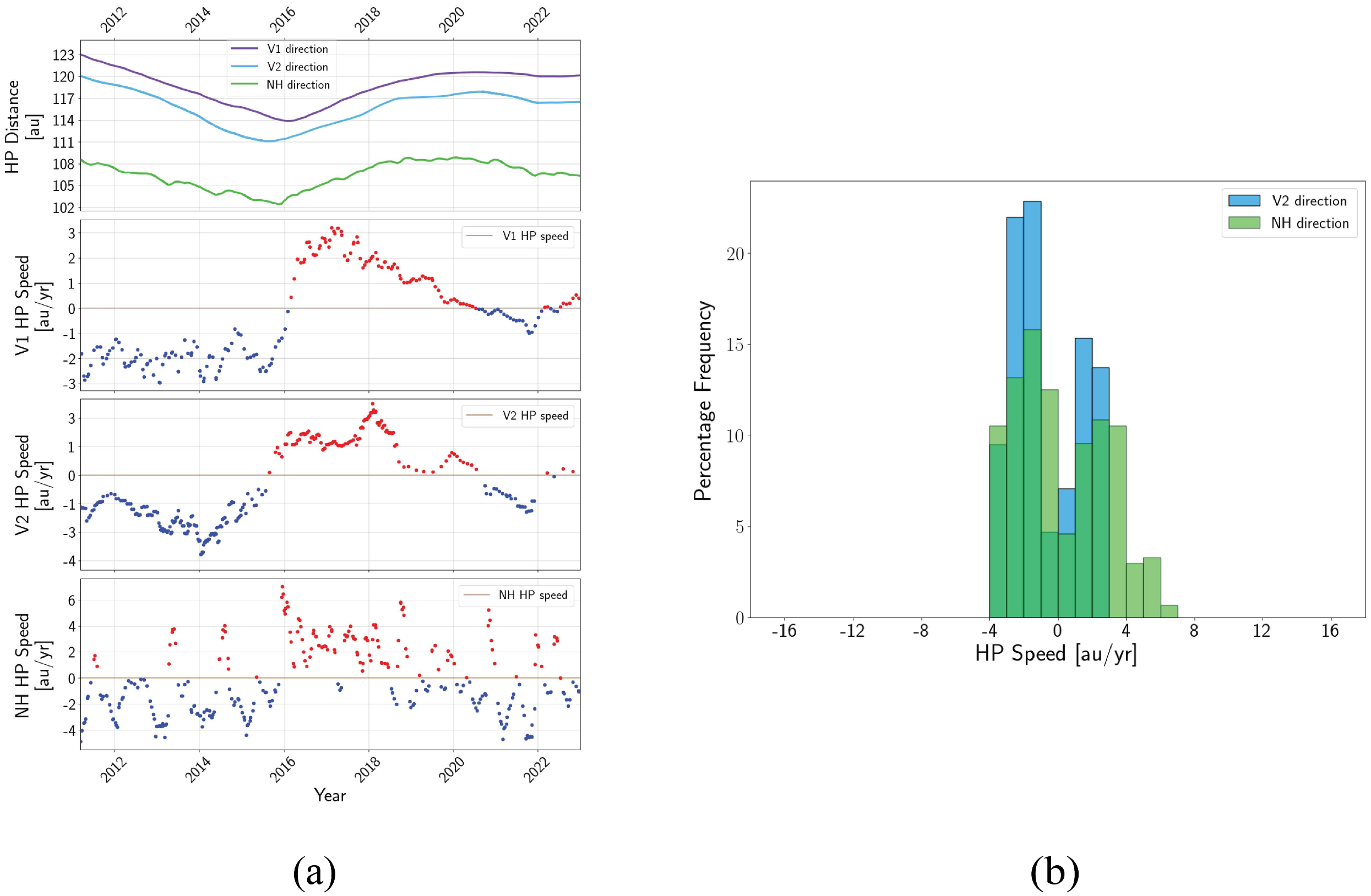}
    \caption{Same as Figure \ref{fig:fig7} but at the heliopause. Overall fluctuations of the heliopause are much weaker than at the termination shock as pressure pulse have a weakened effect. The heliopause moves counter to the termination shock due to the delayed arrival of pressure pulses at the heliopause, and also because of tranmsitted pulses that reduce the pressure force acting on the boundary from outside the heliosphere.}
    \label{fig:fig8}
\end{figure}

We repeat the analysis for the heliopause, in Figure \ref{fig:fig8}. Over 2011 to 2023, the heliopause has an average distance of 106.3 au with 6.4 au excursions from minimum to maximum distance in the NH direction. For V1 and V2, we find an average heliopause distance of 118.5 au with 9.3 au total excursion, and 115.4 au with 9.3 au total excursion, respectively (top panel of \ref{fig:fig8}a). Shorter oscillations in the heliopause location over time are less substantial than at the termination shock, fluctuating within 1 au. Reduced variations in the heliopause motion were also reported in other time dependent studies (e.g. \cite{zankmuller2003,washimi2012,provornikova2014,izmo&alex2020}). The second, third, and fouth panels of Figure \ref{fig:fig8}a show how the heliopause maintains relatively moderate average speeds across the solar cycle: 2.41 au yr$^{-1}$ $\pm$ 1.40 au yr$^{-1}$ along NH, 1.58 au yr$^{-1}$ $\pm$ 0.88 au yr$^{-1}$ along V1, and 2.02 au yr$^{-1}$ $\pm$ 0.84 au yr$^{-1}$ along V2. Compared to the termination shock, the heliopause has a limited range of speeds in its distribution (Figure \ref{fig:fig8}b), reflecting the weakened impact of short time scale variations, driven by fast mode waves or plasma pulses, at the heliopause. However, pulses at the heliopause play a unique role because they transmit across the boundary. Alike the model of \citealt{washimi2017}, pressure pulses that drive into the interstellar medium reduce the pressure acting in the direction of the heliopause. At the same time, enhanced thermal pressure that inundates the heliosheath in the declining phase, tips the pressure balance such that the heliopause continues expanding for a prolonged period, the opposite motion of what occurs at the termination shock.

\section{Discussion and Conclusions} \label{sec:4}

Incorporating time-dependent evolution of the solar wind is essential to understanding the dynamic plasma environment of the heliosphere. We presented a time-dependent simulation of the heliosphere, with a single-ion MHD treatment, and showed how data-driven variability in the solar wind at 1 au influences the heliosphere and interstellar medium interaction upwind. We also studied temporal variability in the solar wind and interstellar medium in the Voyager and New Horizons directions, and directly upwind for the first time in the split-tail heliosphere model.

We found that incorporating 27-day averaged solar activity can reproduce fluctuations in velocity, magnetic field, density, and pressure measured by the Voyager spacecraft. Solar-cycle variations cannot reconcile all differences between observations and model results, such as our model's overestimation of the magnetic field strength, and underestimation of the radial and tangential plasma velocity in the heliosheath 15 au ahead of the heliopause. However, these differences highlight the need for models to incorporate magnetic reconnection in the heliosheath. A more detailed study is needed to understand how processes that reduce the magnetic field will affect the plasma flows and magnetic flux conservation.

Solar activity is not only limited to 11-year periodic features; shorter time scale, plasma pulse structures are frequent and long-lasting in the heliosphere. For the first time, we separate fast and slow mode wave components that manifest as plasma pulses in the heliosheath using a linear Riemann variable analysis. Fast and slow mode waves propagate within the heliosheath, and fast mode waves transmit into the local interstellar medium. The existence of these waves can be a source of turbulence in the heliosheath or compressive fluctuations that advect into the interstellar medium \citep{,zank2017,burlaga2018, fraternale2022,zhao2024}. We reproduce pf2 and show that pressure fronts in the interstellar medium are directly related to fast mode wave structures which mediate solar transient propagation outside of the heliosphere. These waves compress the plasma, creating jumps in magnetic field just beyond the heliopause, just as seen in V1 observations \citep{burlaga2021} and associated electron density fluctuations \citep{kurth2023}. We use linear Riemann variables to track the trajectories of waves back to 1au, and we trace the source of pf2 back to 5 pressure pulses that originated at 1 au in late 2014. Additionally, we show that V2 encountered pf2 in the heliosheath, in the form of ram pressure pulses around 2016.

The termination shock and heliopause behaviors are also influenced by short and long scale variations in the solar wind. We show that the termination shock highly variable and disposed primarily toward inward motions. The termination shock has distinguishing features in its structure and in its speed during rising phases of the solar cycle, in comparison to declining phases. Time-dependent modeling can extend the work of \citealt{powell2025} to examine the connectivity of NH to the termination shock with TSP observations, while considering the time-varying behavior of termination shock. The heliopause is much less variable than the termination shock, but it has longer periods of expansion than we find at the termination shock, due to pressure that is partially driven away from the heliopause, into the interstellar medium. We reserve the time-dependent study of the downwind heliosphere for a future work, where we will explore how  solar cycle effects influence the shape of the heliotail in the croissant-like heliosphere model.

\section*{Acknowledgments}
This work was supported by NASA grant 18-DRIVE18\_2-0029, Our Heliospheric Shield, 80NSSC22M0164, and FINESST award 21-HELIO21-0048. The authors would like to thank the resources provided by the NASA High-End Computing Capability (HECC) program and also the control room staff at NASA Ames Research Center for the use of the Pleiades supercomputer. Special thanks to leading co-authors Merav Opher, Erick Powell, and Senbei Du. Their enthusiasm in exploring the avenues of this study were invaluable in the preparation of this manuscript.

\appendix
\label{sec:appendix}
 In the limit where the magnetic field is perpendicular to the plasma flow, we consider the linearized MHD equations, assuming an arbitrary magnetic field, $\mathbf{B}$, with gradient only in the x-direction. We treat the mean velocity as having only an x-component and denote fluctuating quantities with $\delta$. The equations expressed in matrix form are 

\begin{equation}
    \frac{\partial \bm{\omega}}{\partial t} + \bm{A}\frac{\partial \bm{\omega}}{\partial x} = 0
    \label{eqn:1}
\end{equation}

where $\bm{\omega} = (\delta \rho \quad \delta u_{x} \quad \delta u_{y} \quad \delta u_{z} \quad \delta P \quad \delta B_{y} \quad \delta B_{z})^{T}$, and the matrix $\mathbf{A} $ is the following,

\renewcommand{\arraystretch}{0.8}
\setlength{\arraycolsep}{0.8em}
\begin{equation}
    A = 
    \begin{pmatrix}
        u & \rho     & 0 & 0 & 0              & 0 & 0\\
        0 & u        & 0 & 0 & \frac{1}{\rho} & \frac{B_{y}}{4\pi \rho} & \frac{B_{z}}{4\pi \rho}\\
        0 & 0        & u & 0 & 0              & - \frac{B_{x}}{4\pi \rho} & 0\\
        0 & 0        & 0 & u & 0 & 0 & -\frac{B_{z}}{4\pi \rho}\\
        0 & \gamma P & 0 & 0 & u              & 0 & 0\\
        0 & B_{y}    & -B_{x} & 0 & 0              & u & 0\\
        0 & B_{z}    & 0 & -B_{x} & 0              & 0 & u
    \end{pmatrix}
\label{eqn:2}
\end{equation}

Transforming Equation \ref{eqn:1} by multiplying $\mathbf{L}$---an eigenvector of the matrix $\mathbf{A}$---to both sides of the equation defines the linear Riemann variables which we use in the analysis of the magnetosonic waves in the heliosphere (Equation \ref{eqn:3}).

\begin{equation}
     \frac{\partial \bm{L\omega}}{\partial t} + \lambda\frac{\partial \bm{L\omega}}{\partial x} = \frac{\partial \bm{RV}}{\partial t} + \lambda\frac{\partial \bm{RV}}{\partial x}=0
    \label{eqn:3}
\end{equation}

Above, $\bm {RV = L\omega}$, is the Riemann variable which propagates at a characteristic speed, defined by the eigenvalue, $\lambda$. The right hand side of Equation \ref{eqn:3} is zero, meaning the $RV$s are constant along their characteristic curves; these can be called Riemann invariants \citep{whitham2011linear}. We restrict $\mathbf{B}$ to the z-direction only ($B_x = B_y = 0$) to find two eigenvalues, $\lambda_{f\pm} = u \pm V_{ms}$, corresponding to the fast magnetosonic modes that propagate perpendicular to the mean magnetic field, where $V_{ms}$ is the fast magnetosonic speed. The corresponding eigenvector is

\begin{equation}
    L_{f\pm} = \left( 0 \quad 1 \quad 0 \quad 0
\quad \frac{\pm 1}{\rho V_{ms}} \quad 0 \quad \frac{\pm B}{4\pi\rho V_{ms}} \right)
\label{eqn:4}
\end{equation}

and the resulting $RV$s associated with the fast mode waves are

\begin{equation}
    RV_{f\pm} = \delta u_{x} \pm \frac{\delta P}{\rho V_{ms}} \pm \frac{B\delta B_{z}}{4\pi\rho V_{ms}}
    \label{eqn:5}
\end{equation} with subscript $f$ denoting the $RV$ characteristic along the fast mode wave. The $\pm$ sign physically represents the $RV$ along the forward (+) and backward (-) propagating fast mode wave.

We express $RV_{f\pm}$ in a more intuitive form by multiplying both sides of Equation \ref{eqn:5} by $\rho V_{ms}$ and adding the mean fields (i.e. $\rho u_{x}, P, B^{2}/8\pi$) to the fluctuating quantities as shown below

\begin{equation}
    RV_{f\pm} = \rho (u_{x} + \delta u_{x} ) V_{ms} \pm (P + \delta P) \pm \frac{|B + \delta B_{z}|^{2}}{8\pi} = 
    \rho u_{x} V_{ms} \pm P \pm \frac{B^{2}}{8\pi}
    \label{eqn:6}
\end{equation}.

    The slow mode $RV$s can also be determined in a more general case, in which the magnetic field is not restricted to the z-direction. We still assume the gradient is only in the x-direction. Using the same procedure, we solve the eigenvalue problem, considering the slow mode eigenvalue $\lambda_{s} = u \pm V_{s}$, where $V_{s}$ is the slow magnetosonic speed. In the perpendicular propagation limit,  $V_{s} = 0$, and the corresponding slow mode $RV$s are as follows

\begin{equation}
    RV_{s\pm} = \rho C_{s} u_{z}\sqrt{1 + \beta_s} \pm P \mp \frac{B^2}{8\pi}\beta_s
\end{equation}

where $C_s$ is the sound speed and $\beta_s = 4\pi \rho C_s^2 / B_z^2$. $RV_{s+}$ and $RV_{s-}$ follow the same characteristics in spacetime, so we use $RV_{s+}$ for following the analysis of the slow mode magnetosonic waves.

In Section \ref{sec:3}, $RV_{f+}$, $RV_{f-}$, and $RV_{s+}$, are  denoted as $RV_{f}$, $RV_{rf}$, and $RV_{s}$, respectively. The subscripts represent forward propagating fast mode waves ($f$), reflected fast mode waves ($rf$), and slow mode waves ($s$) in the heliosheath. We note that the RVs calculated are not fully invariant because the MHD background is not homogeneous. Solar wind flow is not constant as it approaches the heliopause and diverts, meaning the bulk flow does not remain largely perpendicular to the magnetic field in all regions. As such, the amplitude of the RVs are not constant and the MHD waves have mixed properties that are coupled to each other.

\bibliography{1A_sources}{}
    \bibliographystyle{aasjournal}


\end{document}